\documentclass[twocolumn,tighten]{aastex63}

\received{\today}
\revised{\today}
\accepted{\today}

\usepackage{bm}
\usepackage{amsmath}
\usepackage{hyperref}

\makeatletter
\newcommand*\bigcdot{\mathpalette\bigcdot@{.5}}
\newcommand*\bigcdot@[2]{\mathbin{\vcenter{\hbox{\scalebox{#2}{$\m@th#1\bullet$}}}}}
\makeatother

\shorttitle{Particles \& Self-Gravity I: Spirals}
\shortauthors{Baehr \& Zhu}
\graphicspath{{./}{figures/}}

\begin{document}

\title{Particle Dynamics in 3D Self-gravitating Disks I: Spirals}

\correspondingauthor{Hans Baehr}
\email{hans-paul.baehr@unlv.edu}

\author[0000-0002-0880-8296]{Hans Baehr}
\affil{Department of Physics and Astronomy, University of Nevada, Las Vegas, 4505 South Maryland Parkway, Las Vegas, NV 89154, USA}

\author[0000-0003-3616-6822]{Zhaohuan Zhu}
\affil{Department of Physics and Astronomy, University of Nevada, Las Vegas, 4505 South Maryland Parkway, Las Vegas, NV 89154, USA}




\begin{abstract}
Spiral arms are distinctive features of many circumstellar disks, observed in scattered light, which traces the disk surface, millimeter dust emission, which probes the disk midplane, as well as molecular emission. The two leading explanations for spirals are wakes generated by a massive planet and the density waves excited by disk self-gravity. We use stratified 3D hydrodynamic shearing-box simulations including dust particles and disk self-gravity to investigate how gas and dust spirals in a self-gravitating disk depend on the simulation size, the cooling efficiency, and the aerodynamics properties of particles. We find that opening angles of spirals are universal ($\sim10^o$), and not significantly affected by the size of the computational domain, the cooling time, or the particle size. In simulations with the biggest domain, the spirals in the gaseous disk become slightly more open with a higher cooling efficiency. Small dust follows the gaseous spirals very well, while intermediate-sized dust with dimensionless stopping time $(\mathrm{St})$ close to 1 concentrates to the spirals more and shows stronger spirals. However, large dust with $\mathrm{St} > 1$ also shows spirals, which is different from some previous simulations. We identify that this is due to the gravity from the gas to the dust component. We show that when $\mathrm{St} \gtrsim Q$, the gravitational force from the gaseous spirals to the dust particles becomes stronger than the particles' aerodynamic drag force, so that the gas significantly affects these large particles through gravitational interaction. This has important implications for both spiral observations and planetesimal formation/dynamics.
\end{abstract}


\section{Introduction}
\label{sec:intro}

Protoplanetary disks are observed with a number of features and substructures, from the common rings of recent surveys \citep{Andrews2018a,Long2018} to the less common but equally intriguing spirals \citep{Grady2013,Perez2016,Huang2018,Johnston2020}, vortices \citep{vanderMarel2013,vanderMarel2016} and potential warps \citep{Benisty2015}. Spirals in the gas can be caused by two distinctive mechanisms, either by the wakes produced by the superposition of higher order modes induced by a point source (i.e. a planet) \citep{Zhu2015,Dong2015,Bae2018a,Bae2018b} or by non-axisymmetric modes of gravitational instabilities \citep{Lin1964,Dong2016,Lee2019a}. Distinguishing between these two sometimes relies on the pitch angle of the spiral arms \citep{Pohl2015,Hall2018}, the number of arms \citep{Bae2018a,Bae2018b} or their amplitude \citep{Juhasz2015}. We focus on the pitch angle for the ease in comparing and constraining this parameter with observations. While the number of arms is easier to constrain observationally, most disks only show two which is in line with theoretical expectations for either cause \citep{Bae2018a}. Due in part to observational constraints, determining which causes the spirals in images, however, has proven to be a challenge \citep{Forgan2018,Forgan2018a,Ren2018}.

Observations come in three regimes, those which are observed in light scattered off the optically thick gas surface of the disk \cite[][etc.]{Muto2012,Grady2013,Garufi2013,Stolker2016,Benisty2015,Dong2018}, those which are observed as spirals in the dust emission \citep{Perez2016,Huang2018,Reynolds2020} with ALMA and molecular line emission \citep{Boehler2018,Booth2019d,Huang2020c}. By only observing the surface of the disk, spirals in scattered light images may not indicate the presence of a feature throughout down to the disk midplane. In TW Hydrae for example, there exists an apparent discrepancy between the spirals in gas observations \citep{Teague2019} and the rings seen in the dust continuum \citep{Andrews2016}. On the other hand, although spirals in the dust may probe the midplane of the disk, they may not reflect the spirals in the gaseous disk \citep{Perez2016}. 

Gravitationally unstable disks are commonly characterized through the presence of spiral arms due to the sheared self-gravitating waves generated by the massive disk \citep{Lin1964,Goldreich1965}. These are typically represented by the dominant $m=2$ spiral mode, but higher order spirals are common in simulations depending on disk mass and disk aspect ratio \citep{Cossins2009,Kratter2016}. Comparatively, a spiral formed by a planet will feature a dominant $m=1$ exterior mode, with higher order interior modes, depending on the mass of the perturber \citep{Bae2018a}. Thus a planet that induces multiple spirals are almost always considered to be located further out than the observed spirals.

However, recent images of potentially self-gravitating disks reveal either no signatures of spirals \citep{Andrews2018a} or only faint spirals \citep{Perez2016} and disks with spirals are not conclusively in the regime where self-gravitational effects should be relevant \citep{Huang2018a}. Similarly, most disks which show spiral features also do not have obvious planet candidates \citep{Boehler2018,Dong2018a}, although this may be due to detection limitations of low luminosity planets \citep{Ren2018}.

Thus, in the absence of more robust indicators, the characterization of the pitch angle perhaps makes it possible to disentangle the cause of spiral features. In the case of an embedded planet, the angle of the spiral is influenced by the mass of the planet as well as the viscosity and local temperature \citep{Rafikov2006,Muto2012}. Overall, the pitch angle is greater close-in to the planet and decreases further away from the planet. For studies which delve into the properties of the spirals generated by embedded planets we refer to \citet{Zhu2015,Meru2017,Hall2018,Dong2018,Dong2018,Bae2018b}.

The spiral structure of self-gravitating disks are often studied for their gas dynamics \citep{Cossins2009,Michael2012}, but the effect of self-gravity with respect to solid material is an often neglected part of mesh-based simulations, due in part to the computational expense and because of the minor effect in low-mass disks. Without considering the self-gravity effects from the gas onto the dust and between the dust particles themselves, larger dust species will remain decoupled from the gas and form axisymmetric structures, and this may be the expected outcome in many cases \citep{Cadman2020a}. However, as we will show, when the effects of the gas self-gravity is included on the dust, even large species will form non-axisymmetric structure similar to the gas.

In this paper, we investigate how the simulation parameters of domain size and cooling timescale affect the pitch angle of gas and dust spiral features in local self-gravitating disks. In Section \ref{sec:spiralstructure} we outline the formation of spirals in self-gravitating disks and then describe the numerical setup in Section \ref{sec:model}. We then present the results in Section \ref{sec:results} and discuss the implications and limitations in Section \ref{sec:discussion} before concluding in Section \ref{sec:conclusion}. 

\begin{deluxetable*}{cccccccc}
\tablecaption{Simulation parameters}

\tablehead{\colhead{Simulation}  & \colhead{Grid Cells} & \colhead{Particle number} & \colhead{$\mathrm{St}$} &\colhead{$\epsilon_{0}$} & \colhead{$L_{x}=L_{y}$} & \colhead{$\beta$} & \colhead{Note}}
\startdata
S\_t2\_B   & $512^2 \times 256$  &$1.5\times 10^6$& $0.01,0.1,1,10,100,1000$  & $10^{-2}$ & $(80/\pi) H$&  2 & -\\
S\_t5\_B    & $512^2 \times 256$  &$1.5\times 10^6$& $0.01,0.1,1,10,100,1000$  & $10^{-2}$ & $(80/\pi) H$&  5 & -\\
S\_t10\_B    & $512^2 \times 256$  &$1.5\times 10^6$& $0.01,0.1,1,10,100,1000$  & $10^{-2}$ & $(80/\pi) H$&  10 & -\\
S\_t2\_BB  & $512^2 \times 128$  &$1.5\times 10^6$& $0.01,0.1,1,10,100,1000$  & $10^{-2}$ & $(160/\pi) H$&  2 & -\\
S\_t5\_BB   & $512^2 \times 128$  &$1.5\times 10^6$& $0.01,0.1,1,10,100,1000$  & $10^{-2}$ & $(160/\pi) H$&  5 & -\\
S\_t10\_BB   & $512^2 \times 128$  &$1.5\times 10^6$& $0.01,0.1,1,10,100,1000$  & $10^{-2}$ & $(160/\pi) H$&  10 & -\\
\hline
S\_t5\_B\_lowpsg   & $512^2 \times 256$  &$1.5\times 10^6$& $0.01,0.1,1,10,100,1000$ & $10^{-6}$ & $(80/\pi) H$&  5 & low dust mass\\
S\_t5\_B\_nopsg    & $512^2 \times 256$  &$1.5\times 10^6$& $0.01,0.1,1,10,100,1000$ & $10^{-2}$ & $(80/\pi) H$&  5 & no particle self-gravity\\
S\_t5\_B\_nogi     & $512^2 \times 256$  &$1.5\times 10^6$& $0.01,0.1,1,10,100,1000$ & $10^{-2}$ & $(80/\pi) H$&  5 & no self-gravity, settling test\\
\hline
S\_t20\_B\_hires   & $1024^2 \times 512$  &$1.5\times 10^6$& $0.01,0.1,1,10,100,1000$ & $10^{-2}$ & $(80/\pi) H$&  20 & -\\
S\_t30\_B\_hires    & $1024^2 \times 512$  &$1.5\times 10^6$& $0.01,0.1,1,10,100,1000$ & $10^{-2}$ & $(80/\pi) H$&  30 & -\\
\enddata
\tablecomments{Simulations in this paper and their resolutions in total grid cell number, total particles included, particle sizes in Stokes number $\mathrm{St}$, initial dust-to-gas density ratio $\epsilon_{0}$, simulation domain length $L$ and cooling parameter $\beta$. All simulations have the same $L_{z} = (40/\pi) H$. The simulation with 'low dust mass' means that particles do not contribute to the gravitational potential while still feeling acceleration from the gas potential. When there is no particle self-gravity at all, the particles neither contribute to the potential nor feel the gas potential, i.e. test particles that feel drag force only.}
\label{tab:sims}
\end{deluxetable*}

\section{Theoretical Framework}
\label{sec:spiralstructure}
\subsection{Spiral Structure}

Spiral density perturbations in a disk are, regardless of the source, described in the form of a plane wave under the assumption of being tightly-wound\footnote{Also known as the WKB approximation: the radial space between any consecutive spiral arms is much smaller than the radius of the disk.}, with potential $X$ \citep{Shu2016,Binney2008}
\begin{equation} \label{eq:3dspiralperturbation}
    X(\bm{x},t) = Xe^{i(\bm{k}\cdot\bm{x} -\omega t)},
\end{equation}
which for a simplified 2D potential in terms of the radial and azimuthal wavenumbers, $k$ and $m$ respectively, is written as
\begin{equation} \label{eq:2dspiralperturbation}
    X(r,\theta,t) = Xe^{i(m\theta + kr - \omega t)}.
\end{equation}

In a self-gravitating disk, the density perturbations are described in linear theory by the Lin-Shu dispersion relation \citep{Lin1964} 
\begin{equation} \label{eq:linshudispersion}
    (m\Omega - \omega)^{2} = c_{s}^{2} k^{2} - 2\pi G\Sigma |\bm{k}| + \kappa^{2},
\end{equation}
which relates the frequency of density perturbations $\omega$ to the azimuthal and radial wavenumbers. Gravitational collapse proceeds for high surface densities $\Sigma$ unless prevented by shear and pressure. The shear, given by the epicyclic frequency $\kappa$, simplifies to the orbital frequency $\Omega = \sqrt{GM_{*}/R^{3}}$ for a Keplerian disk at radius $R$ around a central star of mass $M_{*}$. Also supporting against collapse is the thermal pressure, quantified in terms of the sound speed $c_{s}$.

When $m=0$ and axisymmetric collapse is dominant, Equation (\ref{eq:linshudispersion}) produces the familiar Toomre stability criterion for the gravitational collapse of a gaseous disk \citep{Toomre1964}.
\begin{equation}\label{eq:toomre}
Q = \frac{c_{s}\Omega}{\pi G\Sigma} = 1
\end{equation}
However, when $Q>1$ and higher modes of $m$ are not suppressed by a dominant $m=0$ mode,  non-axisymmetric structure becomes prevalent in a gravitoturbulent disk.

Smaller simulation domains are advantageous for the high grid cell counts per length scale, potentially capturing small scale turbulence, but can suffer from inadequate coverage of stabilizing and destabilizing modes. \citet{Booth2019} showed one effect of a small shearing box includes significant fluctuations away from equilibrium gravitoturbulent densities and stresses. This is attributed to the missing large scale destabilizing radial modes which allow shear to begin pulling apart overdensities before collapsing again. If smaller radial wavenumbers $k$ are suppressed, the higher wavenumbers will dominate and may drive a smaller pitch angle $\theta$
\begin{equation} \label{eq:pitchangle}
\tan\theta = \frac{m}{kR}.
\end{equation}
\citet{Yu2019} suggest that the pitch angle depends largely on the fastest growing radial mode
\begin{equation}
k_{\mathrm{crit}} = \frac{\pi G\Sigma}{c_{s}^{2}},
\end{equation}
such that the pitch angle at the critical wavelength $\lambda = 2\pi/k_{\mathrm{crit}}$ is
\begin{equation}\label{eq:yuangle}
\theta\sim \frac{mc_{s}^{2}}{\pi G \Sigma R}\sim \frac{QmH}{R}.
\end{equation}
If we further assume that $Q\sim1$, we have
\begin{equation}
   \theta\sim \frac{mH}{R}. 
\end{equation}
With smaller simulation boxes (smaller $R$), the pitch angle will increase. In a realistic disk with $H/R\sim0.1$, the pitch angle of the critical mode will be $\sim$10$^o$ for the $m=2$ mode.

\begin{figure*}[t]
\centering
\includegraphics[width=0.48\textwidth]{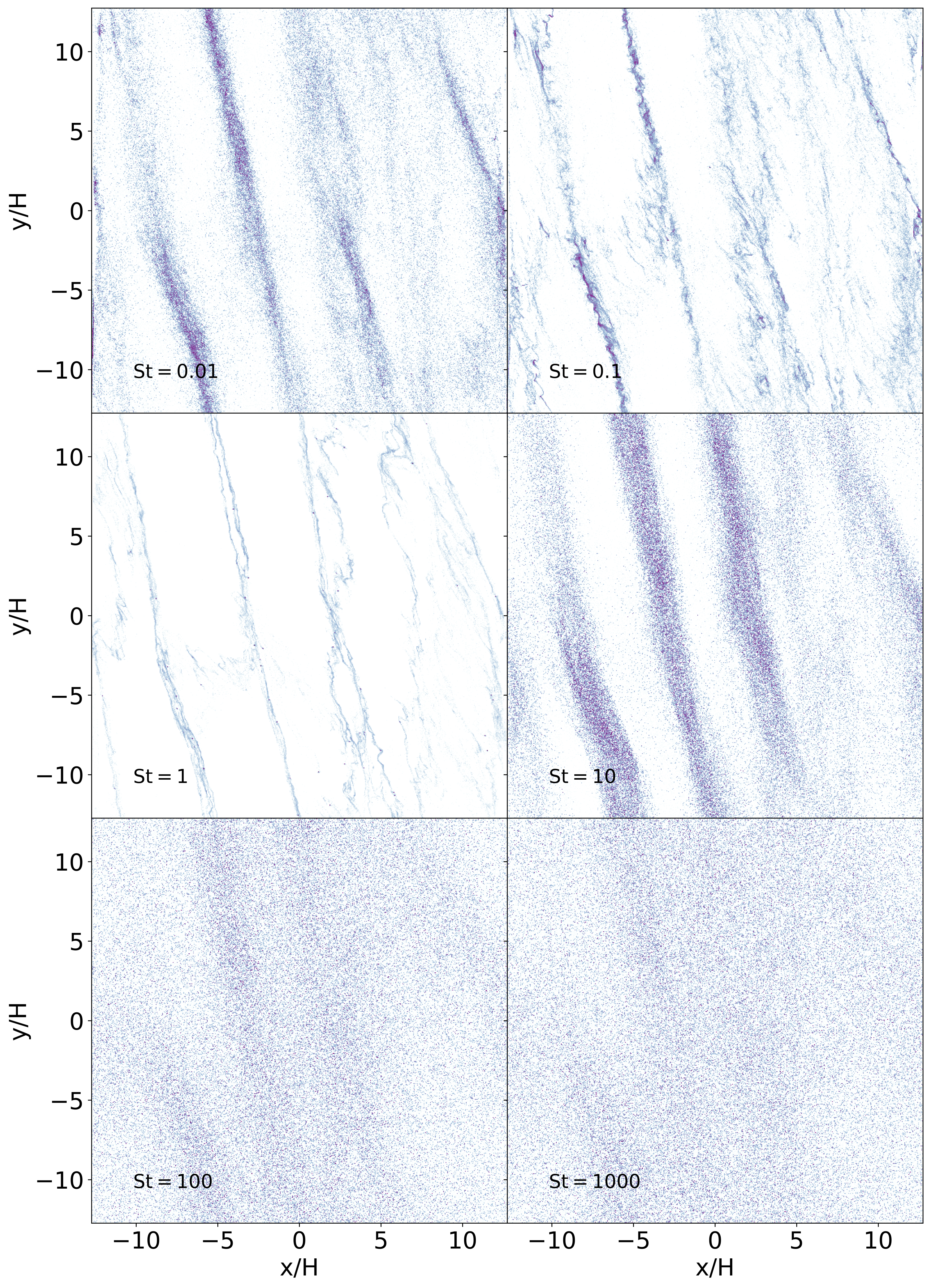}%
\includegraphics[width=0.48\textwidth]{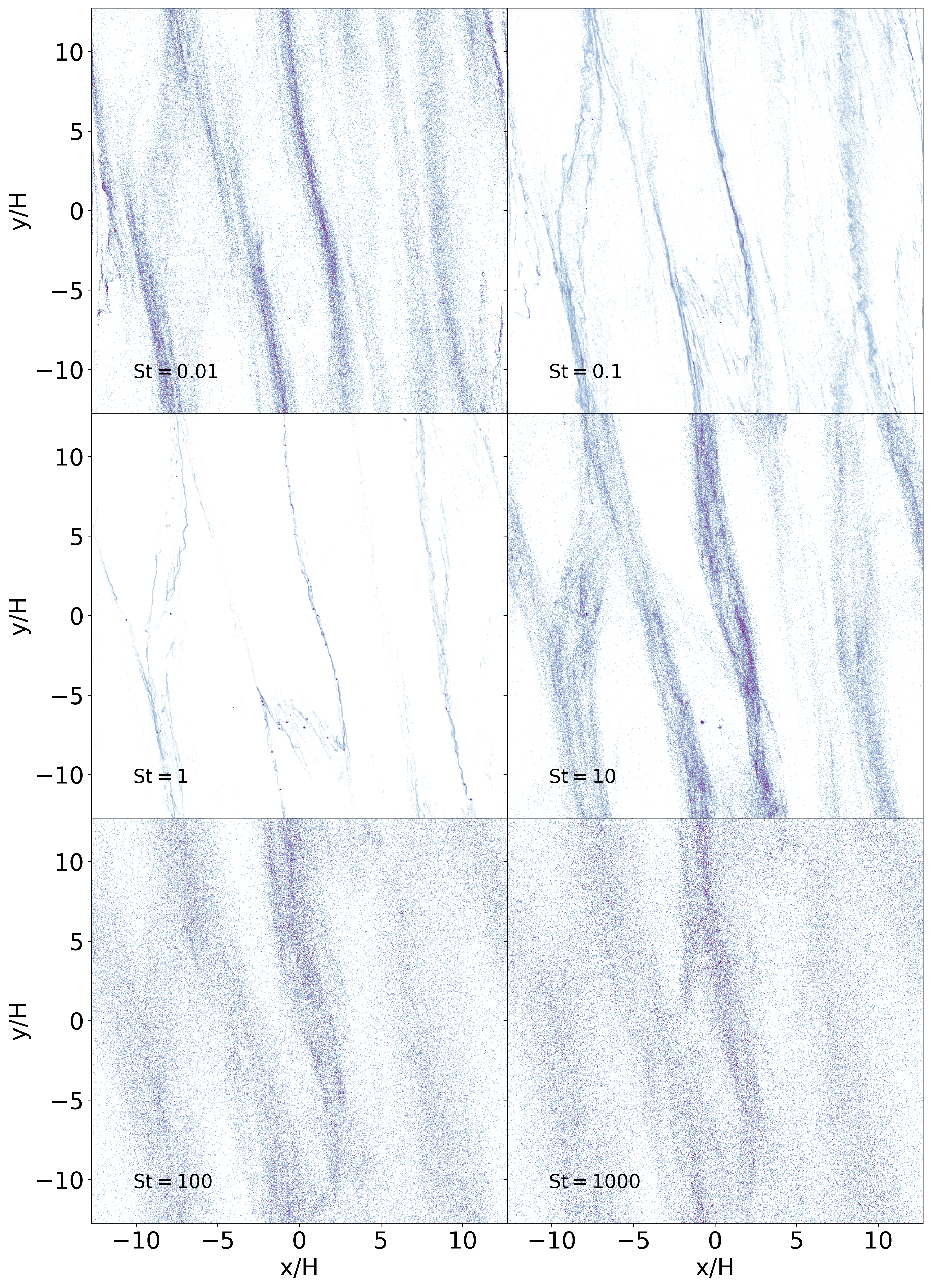}
\caption{Distribution of the various species of particles in the medium-sized simulations for cooling timescales $\beta=2$ (left) and $\beta=10$ (right). While particles were mapped to the grid using a triangular-shaped cloud scheme for calculating drag forces during runtime, here the particles are binned into the single nearest cell.}
\label{fig:dustdensity_species}
\end{figure*}
\begin{figure*}
\centering
\includegraphics[width=0.98\textwidth]{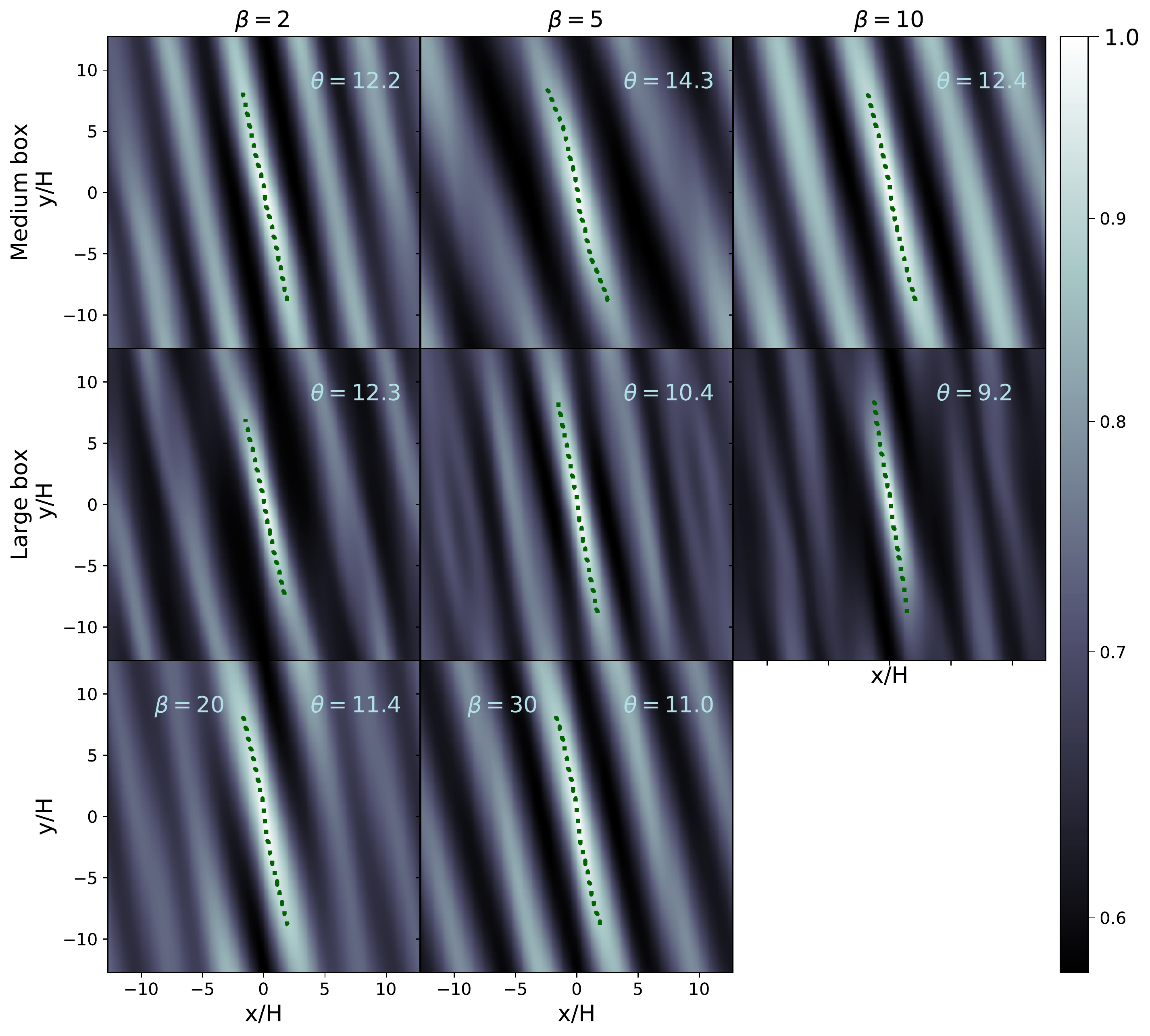}
\caption{The autocorrelation of the gas surface density $\xi$, as calculated from Equation \eqref{eq:autocorrelation}, in the six primary simulations which comprise this study. For each simulation, the autocorrelation was calculated for at the same time, $t=60\,\Omega^{-1}$, when gravitoturbulent features are well-established. Dotted lines indicate the ridge of the autocorrelation maximum, from which a linear fit produces the angle displayed. Additional simulations with longer cooling timescales are included for reference in the bottom row.}
\label{fig:lategaspitchangle}
\end{figure*}
\begin{figure*}
\centering
\includegraphics[width=0.98\textwidth]{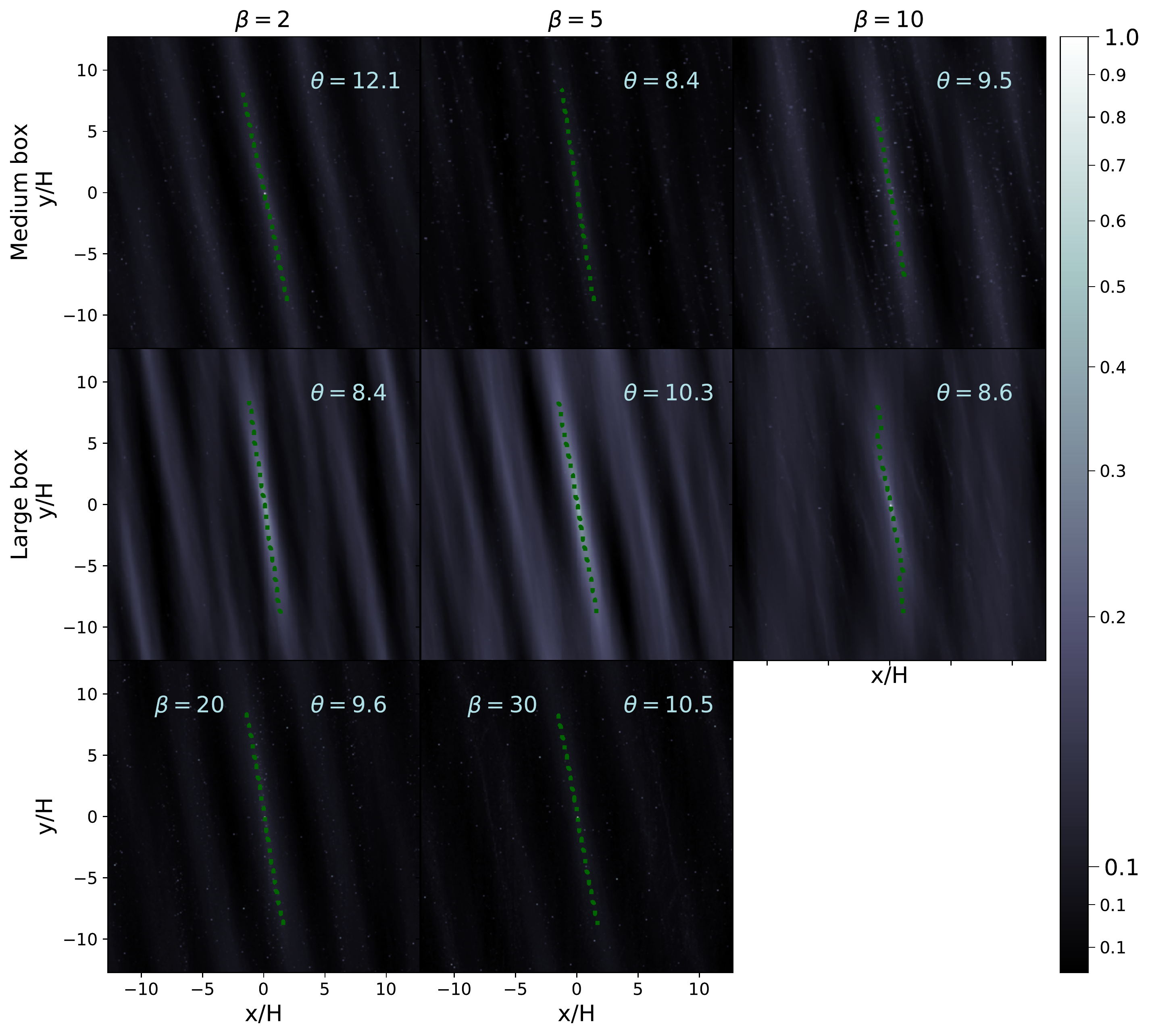}
\caption{The autocorrelation $\xi$ as calculated in Figure \ref{fig:lategaspitchangle}, but for the dust surface density at $t=60\,\Omega^{-1}$.}
\label{fig:latedustpitchangle}
\end{figure*}

Similarly in a shearing box, we can decompose the density perturbation in the Fourier domain with $k_x$ and $k_y$. Then the most unstable mode has 
\begin{equation}
    |k_x|=\frac{\pi G\Sigma}{c_{s}^2}=\frac{1}{HQ}\,,\label{eq:kx}
\end{equation}
which also implies that the simulation domain needs to be larger than $2\pi$H to contain the most unstable mode. With a periodic condition in the $y$-direction, the $m=1$ mode has $k_y=2\pi/L_y$. Thus, the pitch angle is
\begin{equation}
    \theta\sim\frac{k_y}{k_x}=\frac{2c_s^2}{L_y G\Sigma}=\frac{2\pi HQ}{L_y}\,. \label{eq:theta}
\end{equation}

On the other hand, even if the most unstable mode always has $m=1$ or 2, the mode may not represent the spiral in the nonlinear phase. Then, another way to derive the pitch angle is to use the saturated stress in gravitotubulent disks. \cite{Gammie2001} pointed out that cooling leads to disk accretion in gravitotubulent disks so that $\alpha$ is inversely proportional to the cooling timescale ($\beta$). As shown by \cite{Gammie2001}, the $\alpha$ viscosity  due to the gravitational stress in the shearing box is
\begin{equation}
    \alpha_G=\frac{2}{3\langle\Sigma c_{s}^2\rangle} \sum_{k} \frac{\pi G k_x k_y |\Sigma_k|^2}{|k|^3}\,.
\end{equation}
If we assume that one single $k$ component dominates, then we have
\begin{equation}
    \alpha_G=\frac{2\pi G k_x k_y |\Sigma_k|^2}{3\langle\Sigma c_{s}^2\rangle |k|^3}  \sim \frac{2\pi G |\Sigma_k|^2 k_y}{3\langle\Sigma c_{s}^2\rangle |k_x|^2}\,,
\end{equation}
where we have used the fact that $k_x\gg k_y$. If we assume $\Sigma_k\sim f\Sigma$, we can derive
\begin{equation}
    \alpha=\frac{2f^2k_y}{3HQk_x^2}\sim\frac{2f^2\theta}{3HQk_x}\,.\label{eq:theta2}
\end{equation}
If we assume that the most unstable mode (Equation \ref{eq:kx}) still dominates and $f\sim1$, we can further simplify this to $\alpha\sim\theta$. The two estimates above (Equation \ref{eq:theta} and \ref{eq:theta2}) are based on different assumptions and provide very different results about the pitch angle $\theta$. Equation \ref{eq:theta} suggest that the pitch angle depends on the box size while Equation \ref{eq:theta2} implies that the pitch angle depends on the cooling (or $\alpha$). We want to measure the pitch angle from our simulations and compare with these analytical estimates.

Furthermore, the spirals in the dust disk may have different pitch angles from the gaseous spirals, and observations probe both the gas and dust spirals at different wavelengths. Thus, we will look into spirals in both gas and dust disks.

\subsection{Particle Drag and Self-Gravity}
\label{subsec:particleselfgravity}
As will be shown later, the gravity from the gas component to the dust component is crucial for forming the spirals in dust disks. When particles only experience drag forces, smaller species that are more coupled to the gas are expected to closely match the pitch angle of the gas. Larger particles will remain on their initial trajectories for longer and should be largely unaffected by turbulent gas motions. 

Previous work which did not include the gravity from gas to dust \citep[e.g.][]{Gibbons2012,Cadman2020a} suggested that large particles form axisymmetric rings in self-gravitating disks, while in our simulations we will show that large particles still form spirals. Thus, there are several differences between our work and \citet{Gibbons2012}. First, in 2D simulations particles are assumed evenly distributed in the vertical direction, while our vertically stratified 3D simulations show different vertical dust profiles for each particle species \citep{Baehr2021}. Intermediate species settle rapidly, increasing their local concentration and the effect of self-gravity between particles. But secondly, we identify that the gravity from gas to dust is the main cause for the difference.

We compare the relative strengths of the drag and gas gravitational forces with the following simple argument. Considering that GI turbulence is transonic \citep{Gammie2001}, we assume that velocity differences between particles $\bm{w}$ and the gas $\bm{u}$ are on the order of the sound speed $c_{\mathrm{s}}$ for big particles. Then, the acceleration due to the drag force is 
\begin{equation} \label{eq:dragforce}
a_{\mathrm{drag}} = \frac{|\bm{w} - \bm{u}|}{t_{s}} \sim \frac{c_{\mathrm{s}}}{\mathrm{St}\,\Omega^{-1}}.
\end{equation}
To calculate the acceleration due to the gas gravity, we pick a cylinder shape around a dense filament of radius $r$, length $L$, and the excess density $\delta\rho=\rho-\rho_0$ above the background density. The cylinder thus has a side area of $A=2\pi r L$ and  volume of $V=\pi r^{2}L$. With Gauss' law, we have
\begin{equation} \label{eq:gausslaw}
(2\pi r L) a_{\mathrm{grav}} = 4\pi G\delta\rho \pi r^{2}L,
\end{equation}
such that the acceleration due to the filament's gravity is 
\begin{equation} \label{eq:gravityforce}
a_{\mathrm{grav}} = 2\pi G\delta\rho r.
\end{equation}
Spiral filaments typically have radii on the order of $r\approx H$. For a strong spiral whose $\delta\rho\sim\rho_0$, the ratio between $ a_{\mathrm{drag}}$ and $a_{\mathrm{grav}}$ is
\begin{equation} \label{eq:forceratio}
\frac{a_{\mathrm{drag}}}{a_{\mathrm{grav}}} \sim \frac{Q}{\mathrm{St}}.
\end{equation}
For a marginally stable disk with Toomre stability parameter around $Q\sim 1$, this means that gravity shapes the structure for dust as long as $\mathrm{St} \gtrsim 1$. Because self-gravity becomes more important in this regime, large dust that would otherwise be inertial and potentially axisymmetric instead have non-axisymmetric structure.

\section{Model}
\label{sec:model}

We use 3D hydrodynamic shearing box simulations to study the dynamics of self-gravitating disks with Lagrangian super-particles embedded in the Eulerian mesh using the \textsc{Pencil} code \citep{Brandenburg2003}. \textsc{Pencil} is a sixth-order in space and third-order in time finite difference code with MPI for high parallelization. The code also has robust routines for self-gravity and particle-gas interaction and has been thoroughly tested for a number of astrophysical contexts \citep{Youdin2007,Lyra2007,Yang2016}.

For our simulations we require thermal and hydrodynamic properties such that the disk is marginally Toomre stable, i.e. $Q > 1$. That is to say, for the gravitational constant $G$ and $\Omega$ both equal to one, the sound speed $c_{s}$ and combined gas and dust surface density $\Sigma_{\mathrm{0}} = \Sigma_{\mathrm{g,0}} + \Sigma_{\mathrm{p,0}}$ are chosen such that the initial $Q_{\mathrm{0}} = 1.02$ everywhere.

Local simulations allow the Toomre wavelength\footnote{Also known as the largest unstable wavelength} $\sim 2\pi H$ to be well-resolved in the radial and azimuthal directions ($x$ and $y$ in the Cartesian coordinates, respectively) and keep the boundary conditions periodic. The Toomre wavelength needs to be resolved by at least four grid cells to avoid spurious errors growing into larger, unphysical overdensities \citep{Truelove1997,Nelson2006}. We easily satisfy this by resolving each pressure scale height by at least 10 grid cells and thus the Toomre wavelength by around 60 grid cells.

The shearing box simulations used here employ hydrodynamic equations for density, momentum and energy conservation which are linearized and transformed into co-rotating Cartesian coordinates, where $q = -\mathrm{ln}\Omega / \mathrm{ln}R = 3/2$ is the shear parameter:
\begin{eqnarray}
\frac{\partial {\rho_{\mathrm{g}}}}{\partial t} - q\Omega x\frac{\partial {\rho_{\mathrm{g}}}}{\partial y} + \nabla\cdot(\rho_{\mathrm{g}}\bm{u}) &=& f_{D}(\rho_{\mathrm{g}}) \label{eq:finalmassconserve} \\
\frac{\partial \bm{u}}{\partial t} - q\Omega x\frac{\partial \bm{u}}{\partial y} + \bm{u}\cdot\nabla\bm{u} &=& -\frac{\nabla P}{\rho_{\mathrm{g}}} + q\Omega v_{x}\bm{\hat{y}} \nonumber \\ 
 - 2\Omega\times\bm{u} &-& \nabla\Phi + f_{\nu}(\bm{u}) \label{eq:finalmomconserve} \\
\frac{\partial s}{\partial t} - q\Omega x\frac{\partial s}{\partial y} + (\bm{u} \cdot \nabla)s &=& \nonumber \\
\frac{1}{\rho_{\mathrm{g}} T} \Bigl( 2\rho_{\mathrm{g}}\nu\mathbf{S}^{2} &-& \Lambda + f_{\chi}(s) \Bigr). \label{eq:finalenergyconserve}
\end{eqnarray}
In equations (\ref{eq:finalmassconserve}) - (\ref{eq:finalenergyconserve}), $\mathbf{u} = (v_{\mathrm{x}},v_{\mathrm{y}}+q\Omega x,v_{\mathrm{z}})^{T}$ is the gas flow plus shear velocity in the local box, $\rho_{\mathrm{g}}$ is the gas density, $\Phi$ is the gravitational potential of the gas and dust, and $s$ is the gas entropy. The gas and dust are vertically stratified with a sinusiodal gravity profile to avoid a discontinuity at the vertical boundary \citep[cf.][]{Baehr2017}. Viscous heat is generated by the term $2\rho_{\mathrm{g}}\nu\mathbf{S}^{2}$, with rate-of-strain tensor $\mathbf{S}$. Hyperdissipation is applied with the terms $f_{D}(\rho_{\mathrm{g}})$, $f_{\nu}(\bm{u})$, $f_{\chi}(s)$ which for each has the form
\begin{equation} \label{eq:hyperdiff}
f(\xi) = \nu(\nabla^{6}\xi),
\end{equation}
with constant $\nu = 2.5\, H^{6}\Omega$ \citep{Yang2012}. Heat is lost through $\beta$-cooling prescription
\begin{equation} \label{eq:coolingfunction} 
\Lambda = \frac{\rho (c_{\mathrm{s}}^{2} - c_{\mathrm{s,irr}}^{2})}{(\gamma -1) t_{\mathrm{c}}} 
\end{equation}
with $t_{\mathrm{c}}$ given by $t_{\mathrm{c}} = \beta\Omega^{-1}$ and background irradiation term $c_{\mathrm{s,irr}}^{2}$. This background term is set to the initial sound speed so that cooling does not bring the local temperature too close to zero. This cooling prescription has no dependence on the optical depth and thus all regions cool with the same efficiency. In reality, the opacity will be dominated by the small dust grains and an increase of the particle density will increase the local cooling timescale.

Self-gravity of the gas and dust is solved in Fourier space by transforming the density to find the potential at wavenumber $k$ and transforming the solution back into real space. The solution to the Poisson equation in Fourier space at wavenumber $\bm{k} = (k_{x},k_{y},k_{z})$ is
\begin{equation} \label{eq:gravpotential}
\Phi(\bm{k}, t) = -\frac{2\pi G\rho(\bm{k}, t)}{\bm{k}^2},
\end{equation}
where $\Phi = \Phi_{\mathrm{g}} + \Phi_{\mathrm{d}}$ and $\rho = \rho_{\mathrm{g}} + \rho_{\mathrm{d}}$ are the potential and density of the gas plus dust particles combined. This is the default setup for the first six simulations in Table \ref{tab:sims} and we distinguish these six from the next three by the inclusion of gas-particle and particle-particle gravitational interactions self-consistently. Among these three simulations, the simulation with 'low dust mass' indicates that the particle mass is so low that it does not contribute to the gravitational potential $\Phi = \Phi_{\mathrm{g}}$ and thus every particle does not feel the gravity of other particles while still feeling the gravitational acceleration from the gas potential. For the simulation indicated as 'no particle self-gravity', the particles neither contribute to the potential nor feel the gas potential and thus the particles only feel the aerodynamic drag force, the Shearing box inertial forces, and the gravitational force from the central star.

\begin{figure*}[t]
\centering
\includegraphics[width=0.48\textwidth]{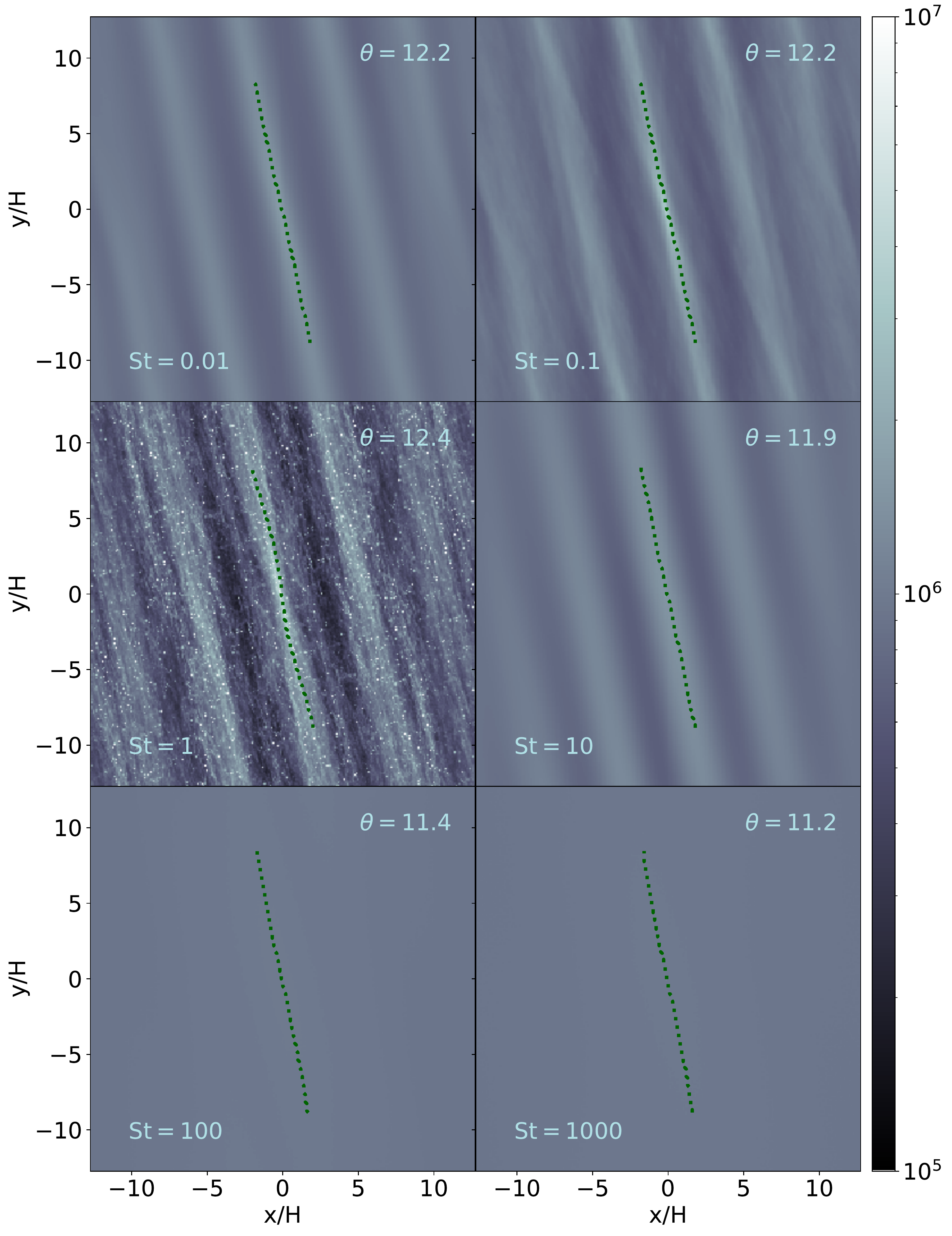}%
\includegraphics[width=0.48\textwidth]{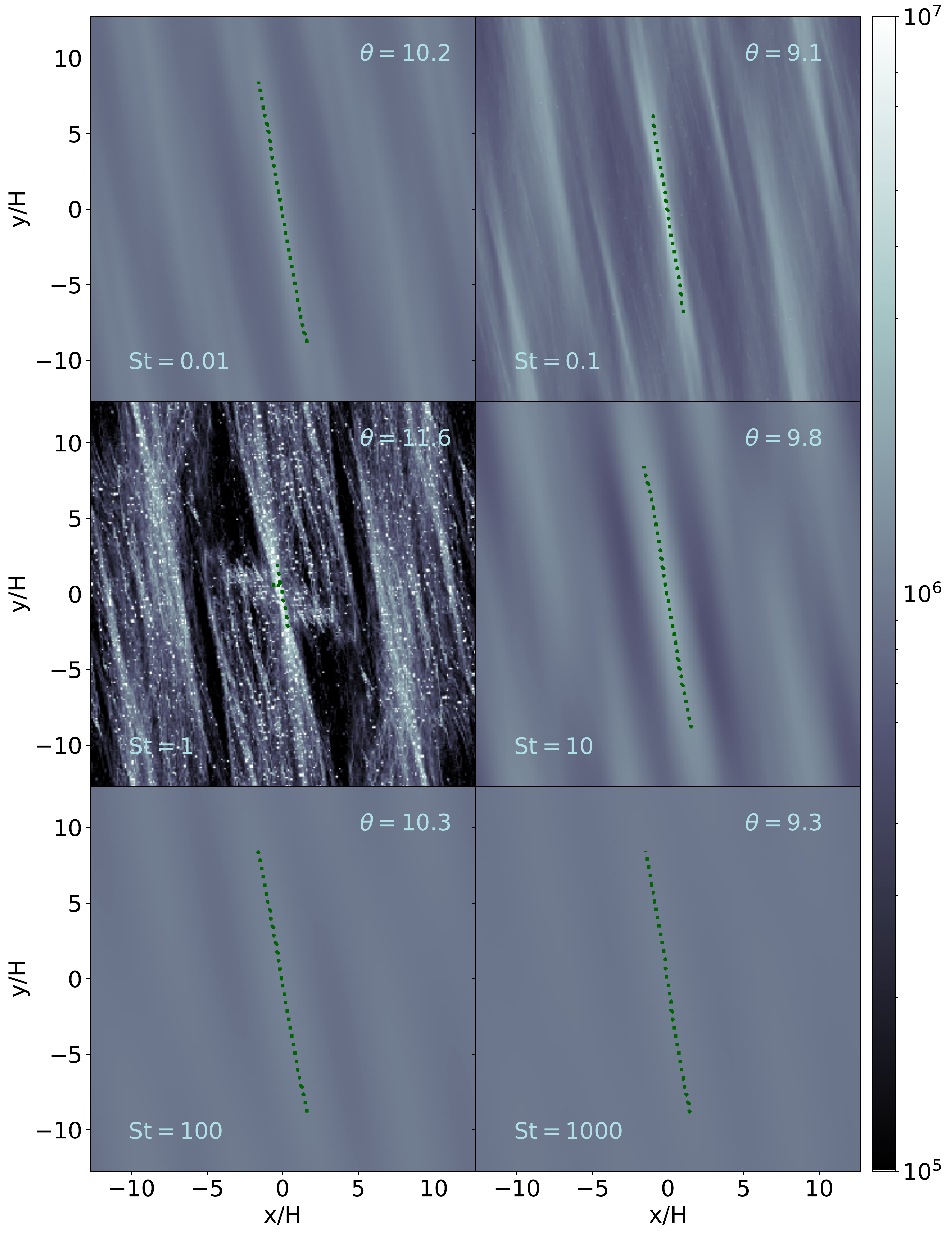}
\caption{Separated autocorrelation calculations $\xi$ for each particle species in simulations with the same medium box simulations from Figure \ref{fig:dustdensity_species}. Particles with $\mathrm{St}=1$ drift towards gas density maxima with the greatest efficiency, creating dense clumps which increase the contrast of the autocorrelation, but also make it more mottled and noisy.}
\label{fig:dustautocorrelation_species}
\end{figure*}

Finally, we use an ideal equation of state, with internal energy $\varepsilon$, and specific heat ratio $\gamma$
\begin{equation} \label{eq:eos}
P = (\gamma - 1)\rho\varepsilon.
\end{equation}
with $\gamma=5/3$.

Particles are initially placed at rest with a vertical Gaussian distribution but are otherwise randomly distributed in $x$ and $y$ directions. The $1.5 \times 10^{6}$ particles are evenly divided amongst six sizes $\mathrm{St} = [0.01,0.1,1,10,100,1000]$ in all simulations. The equations of motion for the particles are then
\begin{equation}
\begin{split}
\frac{d \bm{w}^{(i)}}{d t} &= 2\Omega w_{y}^{(i)} \bm{\hat{x}} - \frac{1}{2} \Omega w_{x}^{(i)} \bm{\hat{y}} - \nabla\Phi \label{eq:particlemotion} \\
&+ \frac{1}{t_{s}} \left( \bm{w}^{(i)} - \bm{u}(\bm{x}^{(i)}) \right)\\
\frac{d \bm{x}^{(i)}}{d t} &= \bm{w}^{(i)} - \frac{3}{2}\Omega x^{(i)} \bm{\hat{y}},
\end{split}
\end{equation}
where $t_{s}$ is the particle stopping time, which we write normalized by the dynamical time $\Omega^{-1}$ as the Stokes number
\begin{equation} \label{eq:stokenum}
\mathrm{St} = t_{s}\Omega.
\end{equation}
In the Epstein regime, the stopping time is related to the particle size as in \citep{Weidenschilling1977}
\begin{equation} \label{eq:epsteinregime}
t_{s} = \frac{a\rho_{\bigcdot}}{c_{\mathrm{s}}\rho_{\mathrm{g}}}
\end{equation}
where $a$ is the particle diameter and $\rho_{\bigcdot}$ is the material density of a dust particle. A super-particle (aka swarm-particle) is a collection of identical dust particles that do not move with respect to the others in the swarm. Particle drag is calculated by interpolating particle positions to the grid using a second-order triangular-shaped cloud scheme \citep[i.e.][]{Yang2018}. Larger particles are less coupled to small scale gas motions, retain their initial perturbations for longer and thus have higher Stokes numbers. On the other hand, smaller particles are well-coupled and will quickly match the velocity of the gas motions in the vicinity and thus have low Stokes numbers. Particle mass is calculated from the gas through the initial condition of the dust-to-gas ratio $\epsilon = \rho_{d}/\rho_{g}$.

Our full collection of simulations and relevant parameters is summarized in Table \ref{tab:sims}. We compare two different box sizes, $L_{x} = L_{y} = (80/\pi) H$ and $(160/\pi) H$, which we label medium and large respectively. Both sizes have the same vertical extent $L_{y} = (40/\pi) H$, but the effective resolution of the larger simulations is half that of the medium ones. Additional simulations at a higher grid resolution were run with $\beta=20$ and $\beta=30$ in order to explore trends to broader disk conditions.

\section{Results}
\label{sec:results}

\begin{deluxetable*}{ccccccccc}
\tablecaption{Measured pitch angles for each simulation}

\tablehead{\colhead{Simulation}  & \colhead{$\theta_{g}$ $[^{\circ}]$} & \colhead{$\theta_{d}$ $[^{\circ}]$}& \colhead{$\theta_{0.01}$ $[^{\circ}]$} & \colhead{$\theta_{0.1}$ $[^{\circ}]$} & \colhead{$\theta_{1}$ $[^{\circ}]$} & \colhead{$\theta_{10}$ $[^{\circ}]$} & \colhead{$\theta_{100}$ $[^{\circ}]$} & \colhead{$\theta_{1000}$ $[^{\circ}]$}}
\startdata
S\_t2\_B   & $12.2$  & $12.1$ &  $12.2$  & $12.2$  & $12.4$  & $11.9$ & $11.4$ & $11.2$\\
S\_t5\_B   & $14.3$  & $8.4$  &  $9.2$   & $8.0$   & $6.8$   & $10.6$ & $12.5$ & $12.7$\\
S\_t10\_B  & $12.4$  & $9.5$  &  $10.4$  & $9.1$   & $11.6$  & $9.8$  & $10.3$ & $9.3$ \\
S\_t2\_BB  & $12.3$  & $8.4$  &  $10.2$  & $9.6$   & $7.4$   & $10.4$ & $8.3$  & $4.8$ \\
S\_t5\_BB  & $10.4$  & $10.3$ &  $9.5$   & $9.7$   & $10.2$  & $10.2$ & $9.9$  & $8.9$ \\
S\_t10\_BB & $9.2$   & $8.6$  &  $6.4$   & $5.9$   & $6.2$   & $8.3$  & $14.1$ & $14.4$\\
\hline
S\_t5\_B\_lowpsg  & $11.5$ & $10.4$ & $10.9$ & $10.4$  & $10.6$ & $10.6$ & $10.6$ & $10.4$ \\
S\_t5\_B\_nopsg   & $13.1$ & $3.3$  & $12.3$ & $13.2$ & $3.0$  & $3.7$  & $0.0$  & $0.0$  \\
S\_t5\_B\_nogi    &    -   &    -   &     -  &     -  &     -  &     -  &     -  &     -  \\
\hline
S\_t20\_B\_hires & $11.4$ &  $9.6$ &  $9.2$ &  $9.4$ &  $9.0$ &  $9.7$ &  $9.5$ &  $9.5$ \\
S\_t30\_B\_hires & $11.0$ & $10.5$ & $10.7$ & $10.4$ & $10.5$ & $10.7$ & $10.0$ &  $9.6$ \\
\enddata
\tablecomments{Diagnostics for the spiral features at settled gravitoturbulence, including pitch angle derived from the gas $\theta_{g}$, total dust $\theta_{d}$, and each particle species separately $\theta_{\mathrm{St}}$.}
\label{tab:spiralresults}
\end{deluxetable*}

Figure \ref{fig:dustdensity_species} shows the vertically integrated particle count for each of the particle species included in two of our simulations after gravitoturbulence has been established. All species trace the gas to some extent, matching the same features of the gas with particularly narrow structures in the case of $\mathrm{St}=1$. Since particles of this size should drift most efficiently towards pressure maxima, this is expected. This also compares well with similar simulations in \citet{Gibbons2012}, particularly for particle sizes $\mathrm{St}=1$ and smaller (see their Figure 2). In \citet{Gibbons2012} however, particles of size $\mathrm{St}=10$ and $100$ show some axisymmetric structure, but do not correspond to the gas features, suggesting more inertial behavior. Discrepancies at these larger sizes can be attributed to two differences between how particles are treated. In our primary simulations, we include the effects of self-gravity of and on the particles, which facilitates dust concentration, and the 3D implementation of these simulations which considers sedimentation effects which proceed faster for moderate Stokes number particles, $\mathrm{St}=0.1$ and $\mathrm{St}=1$. The effects of self-gravity will be discussed later in this section, and the effects of settling will be discussed in Paper II.

It is from these particle distributions and the vertically integrated gas that we calculate the pitch angle $\theta$ in each simulation. To determine the average opening angle, we calculate the autocorrelation of the gas and particle surface densities separately \citep[cf. ][]{Gammie2001,Michikoshi2015a}
\begin{equation} \label{eq:autocorrelation}
\xi(\Tilde{x},\Tilde{y}) = \sum_{x}\sum_{y}\Sigma(x,y)\Sigma(x+\Tilde{x},y+\Tilde{y}).
\end{equation}
Thus each point on the autocorrelation map, $(\Tilde{x},\Tilde{y})$ represents a distance away from each physical point $(x,y)$ on the 2D density map, all superimposed. This allows for a characterization of the overall direction and shape of density structures at a given point in time. Calculating the autocorrelation on a $512^2$ grid was computationally expensive, so all density data at that resolution was interpolated to a $256^2$ grid for calculating the autocorrelation over the entire domain. To calculate the autocorrelation when the distance $(\Tilde{x},\Tilde{y})$ crosses a shear periodic boundary at $x = -L_{x}/2, L_{x}/2$ \citep[cf.][]{Hawley1995}, the density is shifted in the azimuthal direction by $\Delta y=(3/2) \Omega x$ so that features are aligned correctly.

From the autocorrelation, the angle with respect to the azimuthal axis was measured via the ridge of maximum $\xi(\bar{x},\bar{y})$ tracing the diagonal feature, which we mark with a dotted line. The particle autocorrelation is typically less smooth, and thus a Gaussian smooth was applied to the data before determining the line which traces the diagonal feature. This dotted line is then fit with a straight line and the angle $\theta$ is defined as that opened counterclockwise with respect to the azimuthal ($y$) axis.

\begin{figure*}[t]
\centering
\includegraphics[width=0.48\textwidth]{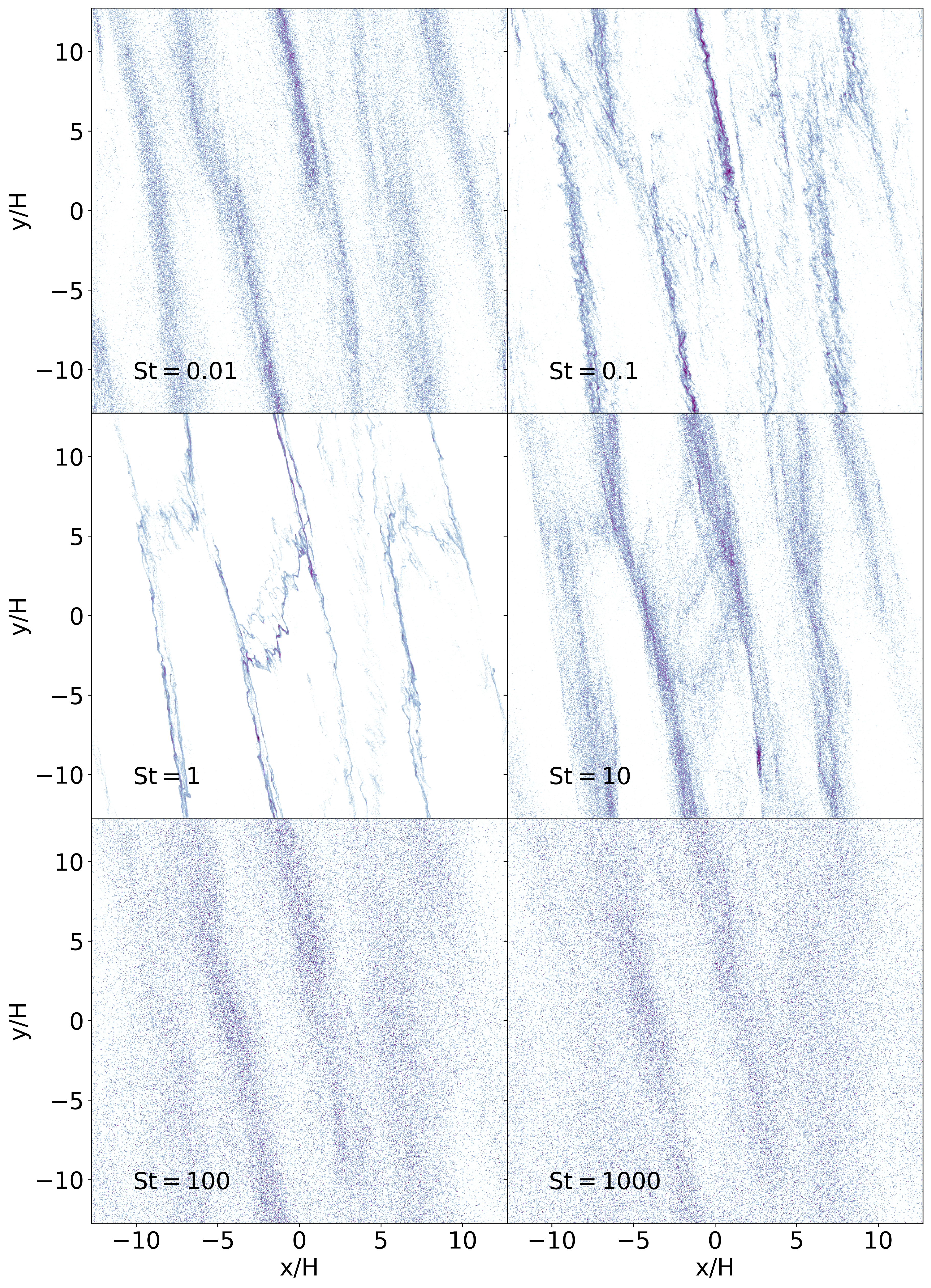}%
\includegraphics[width=0.48\textwidth]{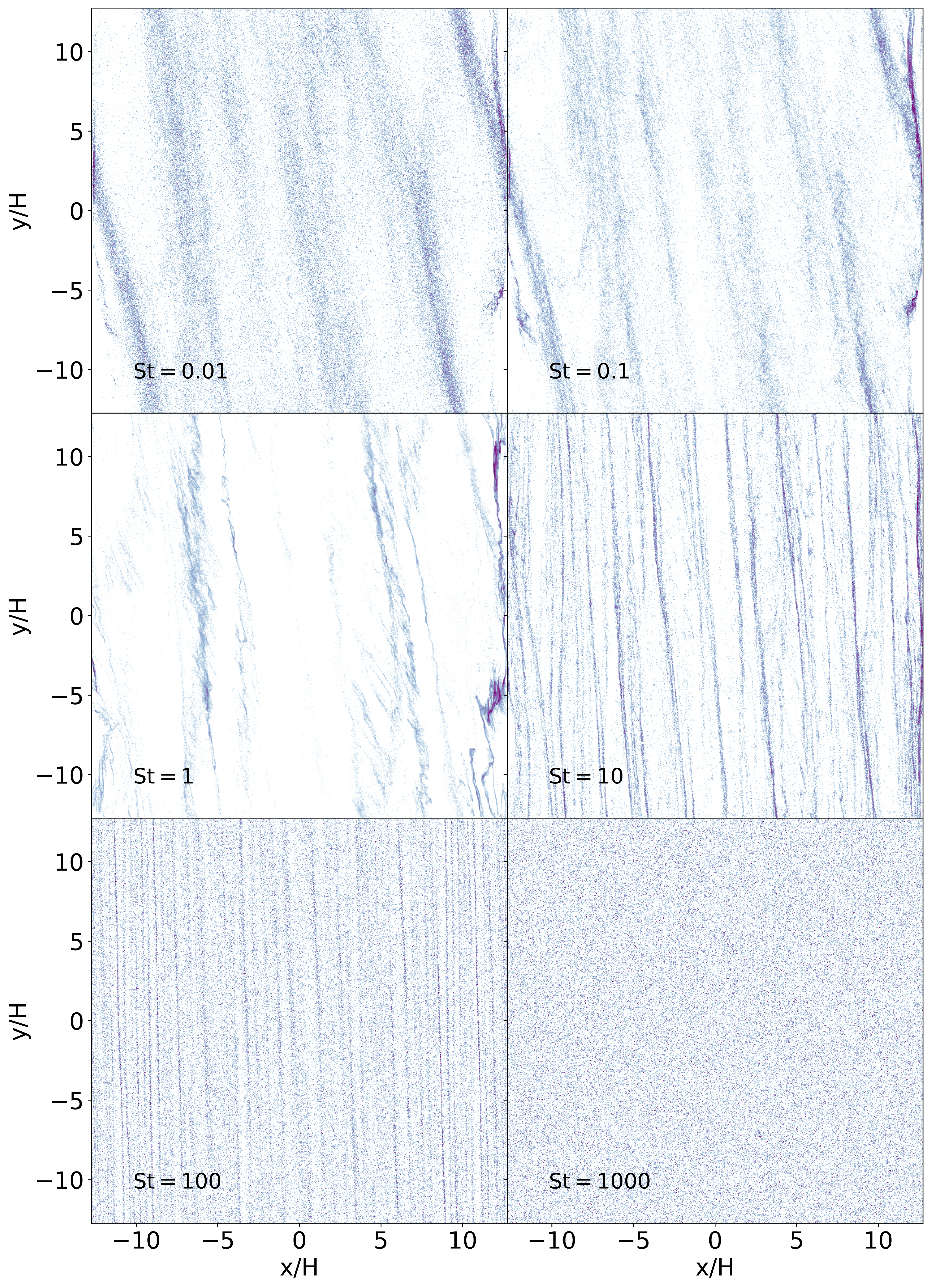}
\caption{Distribution of all particles separated by species in the simulations where particles feel the gas gravity but not each other \texttt{S\_t5\_B\_lowpsg} (left) and where particles do not feel self-gravity from either the gas or dust \texttt{S\_t5\_B\_nopsg} (right), at time $t=60\,\Omega^{-1}$. While in the latter case large particles, $\mathrm{St}=100$ and $\mathrm{St}=1000$, are very inertial with little discernible large scale structure, but when including the effect of the self-gravitating gas, the same particles show more noticeable features.}
\label{fig:dustdistributionpsg}
\end{figure*}

We focus our efforts at a time where gravitoturbulence has been established in the simulation, such that non-axisymmetric (non-linear) features have been allowed to evolve over at least a few dynamical timescales. In Figures \ref{fig:lategaspitchangle} and \ref{fig:latedustpitchangle}, we show the gas and total dust autocorrelation calculations at $t=60\,\Omega^{-1}$. Overall, gas pitch angles $\theta_{g}$ are within a few degrees of $\theta = 12^{\circ}$, with an apparent decreasing trend with increasing $\beta$ for the larger simulations. 

The particle autocorrelation is partially complicated by the dust clumping due to particle drift and particle-particle self-gravity at high particle concentrations. Numerous bright spots make it hard to find the overall dust pitch angle, particularly for $\mathrm{St=1}$. However, there is less clumping at other sizes such that the overall dust picture is not disrupted significantly.

\begin{figure*}[t]
\centering
\includegraphics[width=0.48\textwidth]{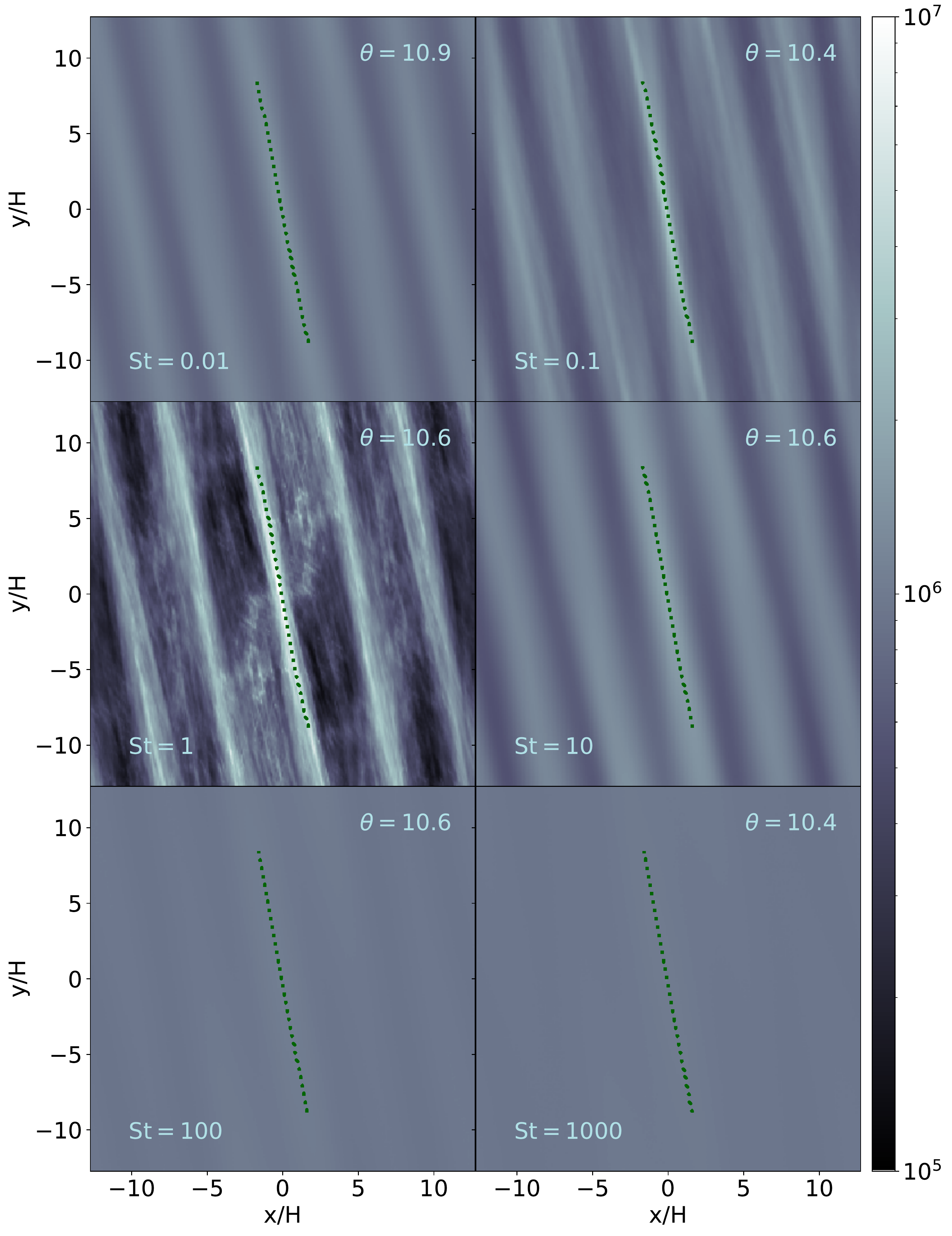}%
\includegraphics[width=0.48\textwidth]{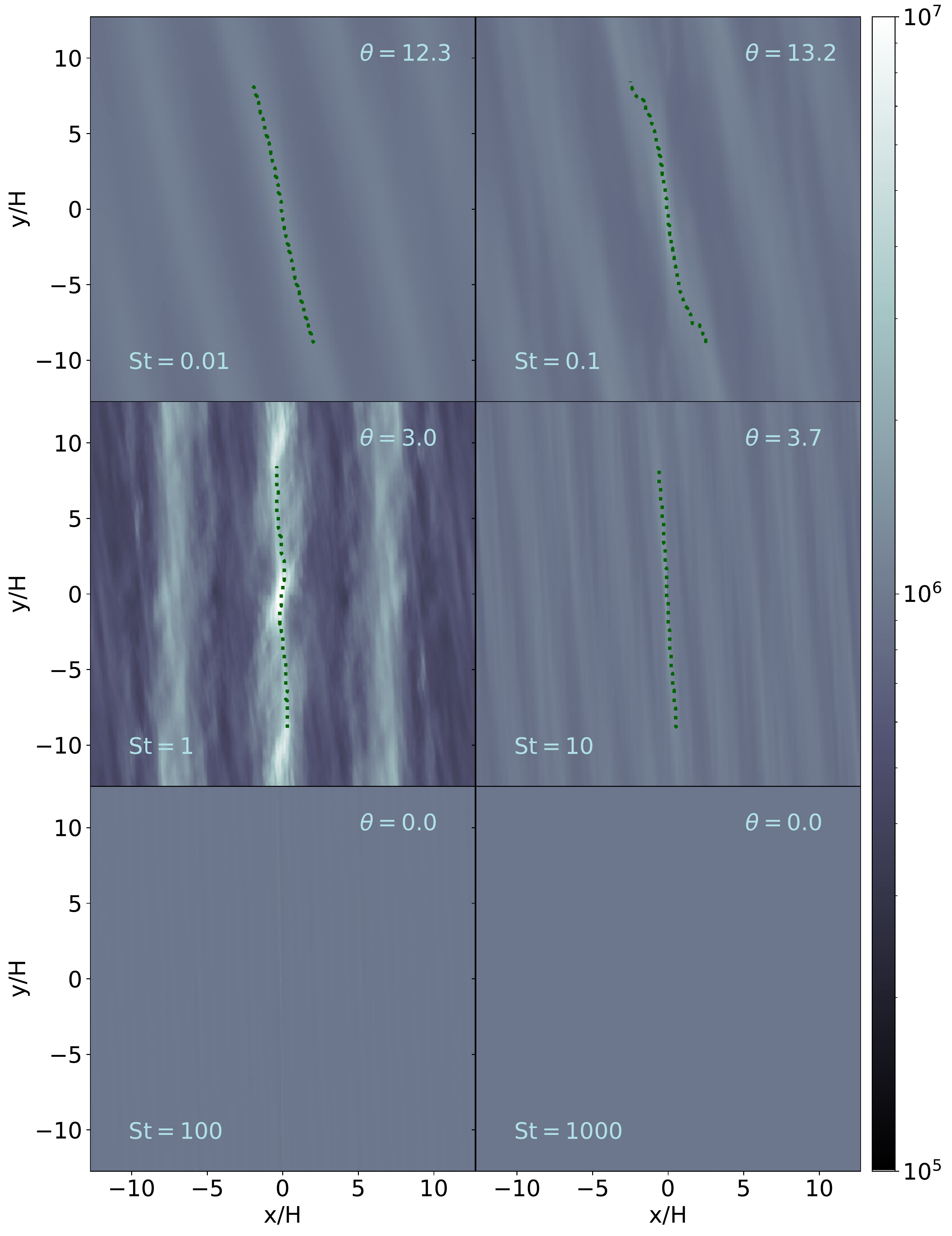}
\caption{Autocorrelation maps for two medium simulations with $\beta = 5$. These are identical to the simulation \texttt{S\_t5\_B} except the left panel excludes particle-particle self-gravity effects (while keeping gas-to-particle gravity) and the right panel excludes all self-gravity effects on individual particles. Even without self-gravity between particles, the pitch angle is consistent over multiple dust sizes. When particles only feel drag, spiral pitch angle declines to $\theta = 0$ with increasing Stokes number $\mathrm{St}$.}
\label{fig:dustautocorrelation_species_psg}
\end{figure*}

We further break down the autocorrelation analysis by particle size. Figure \ref{fig:dustautocorrelation_species} shows two medium-sized simulations at $t=60\,\Omega^{-1}$. Particles with size $\mathrm{St}=1$ are especially prone to clumping, such that densities along filaments are higher and the autocorrelation shows a starker contrast. This makes determining the pitch angle at this species more difficult than the rest and this is reflected in the larger variation of pitch angles at this size. Overall, the spirals shown by different sized particles have surprisingly consistent pitch angle. The pitch angle is almost unchanged even if the $\mathrm{St}$ of the dust changes by 5 orders of magnitude.

In Figure \ref{fig:dustdistributionpsg}, we plot the radial-azimuthal distribution of particles from two simulations identical to our initial six simulations, aside from how particle self-gravity is treated. The left panel includes particle density as part of the potential, but the particles only have $10^{-4}$ times the mass. Thus, they contribute little to the overall potential, even at high local concentrations. However, the particles still feel the acceleration of the gas potential and thus are still affected by dense gas structures. As the autocorrelations show in the left side of Figure \ref{fig:dustautocorrelation_species_psg}, particles still retain effectively the same non-axisymmetric structure at all sizes, although it does not perfectly correlate to the gas. 

In the other case, shown in the right panel of Figure \ref{fig:dustdistributionpsg}, particle density is not included in calculating the self-gravity potential and the acceleration of the potential is not included in Equation \eqref{eq:particlemotion}. In this situation, particle-gas interaction is through the aerodynamic drag only. Thus, larger dust particles become increasingly axisymmetric with increasing dust size ($\theta$ approaching $0^{\circ}$), until they are randomly distributed at $\mathrm{St}=1000$. These axisymmetric features are similar to the results shown in \cite{Gibbons2012}. On the other hand, the left panels in Figure \ref{fig:dustdistributionpsg} are consistent with previous 2D simulations by \cite{Shi2016} who have shown spiral features for large particles. By comparing the left and right panels of Figure \ref{fig:dustdistributionpsg} and \ref{fig:dustautocorrelation_species_psg}, we identify that non-axisymmetric structure can occur for $\mathrm{St}>1$ since the self-gravity of the gas is acting on the dust.

\section{Discussion}
\label{sec:discussion}

\subsection{Gas and Dust Spirals}

The pitch angles measured in all simulations are provided in Table \ref{tab:spiralresults}. The pitch angles of the gaseous spirals are $\sim$10$^{\circ}$ universally and vary by a degree or two over time. For shearing box simulations, the most unstable mode as in Equation \ref{eq:theta} has $\theta\sim14^{\circ}$ using $L_y=(80/\pi) H$ (equivalent to $H/R=0.25$ in a global disk) for our mid-sized box and $m=1$ shown in Figure \ref{fig:dustdensity_species}. Although this is close to $10^{\circ}$, it cannot explain the similar $10^{\circ}$ pitch angles observed in large box simulations with $L_y=(160/\pi) H$ (equivalent to $H/R=0.12$ in global disks) and $m=1$.

Furthermore, for these large box simulations, we notice that the pitch angles increase slightly with a faster cooling time (smaller $\beta$), which is more consistent with Equation \ref{eq:theta2}. For a disk with a faster cooling and a larger $\alpha$, the spirals need to be more open so that the gravitational interaction is more efficient to transport angular momentum. On the other hand, we didn't observe significant change of the pitch angle (e.g. there is no factor of 5 change with $\beta$ changing from 10 to 2), which may suggest that the $f$ factor in Equation \ref{eq:theta2} also depends on the cooling rate (e.g. a faster cooling leads to a stronger spiral so that $f$ is closer to 1). Our simulations shed light on the gaseous spiral structures, but the explanation still deserves more analytical and numerical calculations in future. 

The overall dust structure generally matches the gas well. In most cases, the pitch angles of the dust spirals are within a few degrees below the gas pitch angle. We find that the most surprising feature is the weak or complete lack of a decreasing pitch angle with an increasing particle size. If particles are dominated by their aerodynamic properties (gas-to-particle drag forces only), one might expect only the more well-coupled species to match the gas and decoupled particles to have more axisymmetric structure \citep{Gibbons2012,Cadman2020a}. Two-dimensional simulations at longer cooling timescales have shown decreasing pitch angle \citep{Shi2016}, but we do not observe a considerable decrease in $\theta$ with increasing $\beta$. We occasionally see some very low pitch angles in the dust despite a typical gas pitch angle. We attribute this to interactions between the spiral arms that temporarily disrupt the dust structure but the dust later returns to tight agreement with the gas structure.

\begin{figure}[t]
\centering
\includegraphics[width=0.48\textwidth]{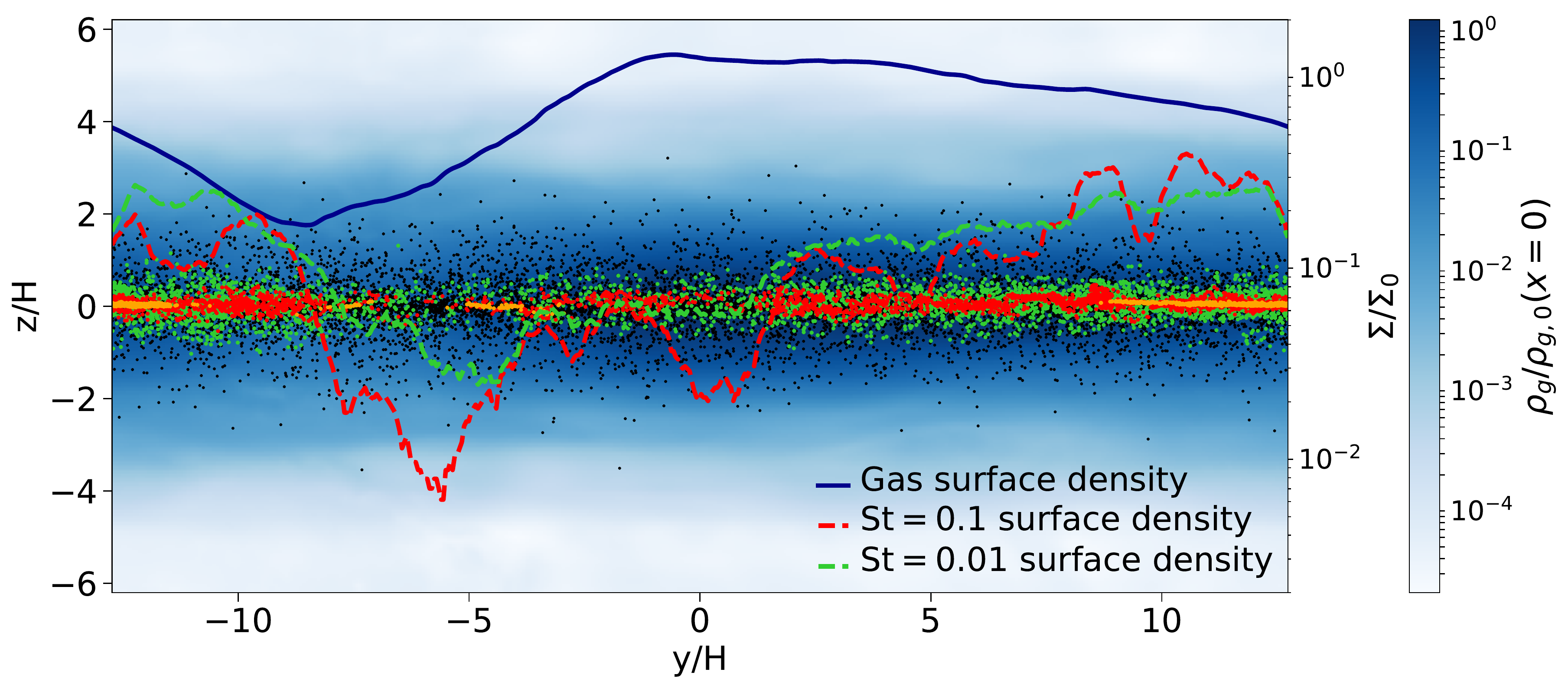}
\caption{A slice of constant radial distance $(x=0)$ showing the particle and gas distribution. Highlighted species are $\mathrm{St}=0.01$ (green), $\mathrm{St}=0.1$ (red), and $\mathrm{St}=1$ (orange) with all other species as black points. The variation in the surface density of the two smaller sizes is indicated by a dashed line with corresponding color.}
\label{fig:azimuthalslice}
\end{figure}

Even if spirals are a ubiquitous feature among all particle species, it may be possible that the contrast between regions of high and low dust concentrations are not discernible as spirals at all. In Figure \ref{fig:azimuthalslice}, we look at a slice of a simulation at constant radius $(x=0)$ and compare the dust surface density of the smaller species that might be observable. That the surface densities of the small dust varies by at least an order of magnitude suggests that these spiral features would not be confused with rings in observations.

We note a slight decrease in the pitch angle with particle size in the simulation \texttt{S\_t2\_BB}, but the pitch angle does not decrease to $\theta = 0$ as it does for purely aerodynamic particles and still has some non-axisymmetric structures. The situation is reversed in \texttt{S\_t10\_BB}, and the larger species have significantly larger pitch angles than the other species and even the gas. Gas pitch angles are not constant in time and for both cases, this could be a result of a delayed response to a temporary shift in the gas structure to a smaller or larger pitch angle. Thus, the smaller dust would more quickly match the angle of the gas through gas drag and the larger dust would require more time. Similar disparities between gas and magnetic field structures have been noted in \cite{Guan2009}.

\citet{Yu2019} analyzed the pitch angles of a number of potentially self-gravitating disks with spirals and find a dependence of the angle on disk mass through the Toomre parameter. Many of the disks they analyze with $Q<4$ have pitch angles from $7^{\circ}$ to $15^{\circ}$ and larger angles for larger values $Q$. Our simulations with a $Q=1$ disk mostly fall within the range of $\theta = 10 - 12 ^{\circ}$. We, however, do not explore how the structure in our simulations varied with initial Toomre $Q$ and leave that for future investigations.

\subsection{Particle Size}
\label{subsec:particlesize}

Particles of size $\mathrm{St}=0.01$ and $\mathrm{St}=0.1$ are the most well-coupled and similar to the cross-correlation of dust and gas in \citet{Riols2020}, within the gas structure. While they will concentrate along gas structures, coupling to small scale gas motions will prevent the formation of narrow, concentrated filaments. This does not mean that small particles necessarily match the pitch angle of the gas the best. 

This is corroborated by the intermediate size $\mathrm{St}=1$, the size which drifts especially well and thus concentrate efficiently into narrow structures. When particle-particle self-gravity is included, these concentrations are able to collapse into denser clumps of particles as long as the local particle density is high enough. The collapse of dense particle clouds can be modeled similarly to unstable gas \citep{Gerbig2020}, resulting an unstable $m=0$ dust mode and occasional strong local axisymmetric structure ($\theta = 0^{\circ}$) in the dust. Only for this size do we see particles concentrate enough for the dust-to-gas ratio to reach unity \citep{Baehr2021}. This is the point where dust backreaction, had it been included in our simulations, becomes relevant. With backreaction, dust concentration will become more pronounced, and dust may concentrate for a broader range of particle sizes. Pitch angles at this dust size show the most variability across all simulations, just as often in the range $\theta = 6 - 8 ^{\circ}$ as in the overall average of $\theta = 10 - 12 ^{\circ}$.

Even larger particles, which should be much less affected by turbulent gas flow, have remarkably similar structures especially when comparing the two largest particle sizes in the right-hand plot of Figure \ref{fig:dustdistributionpsg}. This already hints at something other than aerodynamic drag molding the particle structure at these large sizes and we identify as the gravitational potential of the gas disk.

This is particularly noticeable in the $\mathrm{St}=100$ and $\mathrm{St}=1000$ cases on the left half of Figure \ref{fig:dustdistributionpsg}, where both species are very similar in overall structure despite having very low dust mass and thus even considerable concentrations of particles do not affect the gravitational potential. This is very similar to the more massive particle cases in the plots of Figure \ref{fig:dustdensity_species}, except that particle clumping does not occur at $\mathrm{St} = 1$, which appears to be the largest difference between the two simulations with drastically different particle masses.

While the disk cooling timescale does not have a strong overall impact on the pitch angle of spiral features, it does have an effect on how concentrated features are for $\mathrm{St}>1$ (comparing the left and right panels in Figure \ref{fig:dustdensity_species}). The density concentration for $\mathrm{St}>1$  particles within filament is stronger with a longer cooling timescale. Simulations including self-gravitating particles in 2D simulations by \citet{Shi2016} show the same characteristics. They note that since turbulent strength decreases with less efficient cooling, i.e. $\alpha = (4/9)(\gamma (\gamma - 1)\beta)^{-1}$ \citep{Gammie2001}, particles are stirred less vigorously by the gas and thus relative velocities between particles decrease by an order of a few. With lower particle velocities, gravitational forces from the gas structures begin to have an effect on particle trajectories.

Thus, based on the simulations presented, we suggest Equation \eqref{eq:forceratio} serves to determine when particles of a particular size form non-axisymmetric structures due to aerodynamic drag interactions with the gas or via gravitational interaction with the gas.

\subsection{Box Dimensions}
\label{subsec:boxdimensions}

Shearing boxes are a useful approximation for their high resolution of predominantly local phenomena, but become unreliable when the simulation domain is too small to resolve all or even some unstable wavenumbers \citep{Booth2019} or when the domain is so large that the local linear approximation no longer remains tenable. The former is largely problem dependant and what may be too small for gravitational instabilities is still suitable for simulating magnetorotational instabilities. While the large and medium simulations used here are suitable for calculating spiral pitch angles, anything smaller does not develop sufficient gravitoturbulence to form the necessary large scale structures.

\citet{Hirose2017} calculated autocorrelation maps for various domain sizes and concluded that their smallest boxes do not locally transport angular momentum as well as larger boxes. This also agrees with the study of \citet{Booth2019} which showed that small shearing boxes produced erratic behavior once stabilizing radial modes did not fit within the domain. This causes large swings in the density fluctuations and gravitoturbulent stress which would adversely affect the measured pitch angle. Ultimately, similar to the comparable 2D simulations of \citet{Shi2016}, we find gas and dust structures do not differ greatly between the two domain sizes studied here.

\subsection{Spirals Generated by Planets}
\label{subsec:planetspirals}

Spiral arms can also be attributed to an embedded planet, caused by the superposition of multiple modes of azimuthal and radial disturbances in the gravitational potential which constructively and destructively interfere with each other to produce the spiral morphology seen in simulations \citep{Goldreich1979,Bae2018a,Miranda2019}.

Because these spirals are launched from the planet location they have a very steep pitch angle that becomes shallower the further away from the planet it propagates and will not have a single pitch angle. Thus, the spirals from planets will have a range of pitch angles between $0^{\circ}$ and $20^{\circ}$, depending on the location away from the planet, but not vary significantly with time \citep{Bae2018a}. This differs from the spirals formed by gravitational instabilities, which may vary with time, but have a generally consistent pitch angle over the range of the disk for any given snapshot in time.

Planets have been proposed as progenitors of spirals in many disks, including the frequently studied disks SAO 206462 \citep{Muto2012,Grady2013,Dong2015}, MWC 758 \citep{Dong2015,Dong2018,Boehler2018}, and HD 100453 \citep{Benisty2017,Wagner2018}. The spirals in these disks are unlikely to be caused by gravitational instability in large part due to their older ages and lower disk masses, none being in the ranges normally associated with a self-gravitating disk. Even so, the absence of a detected companion keeps open the possibility of gravitational instabilities as a cause.

\section{Conclusion}
\label{sec:conclusion}

We investigated the spiral arms in self-gravitating disks and analyzed a number of factors which could affect what form they take and how they are observed. Using local 3D shearing box simulations of gravitoturbulent disks we focused on the influence of the gas cooling rate and simulation domain and how different sized particles presented in each case. Self-gravitating disks are a possible cause of spiral features which have been observed in both gas and dust disks.

In our simulations, We characterize their features in not only gas properties, but in the dust as well, which allows for a better comparison with observations. We summarize our results in the following points:

\begin{itemize}
    \item The opening angles of gas spirals in GI disks are universal ($\sim10^o$), which are not significantly affected by the size of the computational domain. In our simulations with the biggest domain size, the gas spirals become slightly more open with increased cooling efficiency.
    
    \item Particles of different sizes have similar responses to the gas structure as measured by the pitch angle when gas self-gravity acts on the dust. While it is expected that smaller particles are more coupled and have pitch angles that are more similar to the gas, we find that this is also true for the larger particles which are less affected by aerodynamic drag but more affected by the gas gravity.
    
    \item The above point suggests drag and self-gravity may be comparable influences on the shape spirals take. When particles do not feel the potential of the gas at all, the ability of large particles to form features corresponding to the gas is drastically diminished. This suggests that self-gravity plays a role in shaping the structure spirals in gravitoturbulent disks, particularly for larger particles. Even very low mass particles behave very similar to more massive particles, but will not collapse into small dense clouds unless particle-particle self-gravity is included.
    
    \item While large particle species show very broad non-axisymmetric features, they become more refined with decreased cooling efficiency (increasing $\beta$). This occurs even for a vertically unsettled distribution of particles which are still undergoing overdamped oscillations about the midplane.
\end{itemize}

Overall, both gas and dust at different sizes show spirals with $\sim 10^{\circ}$ pitch angle (almost all from 6$^{\circ}$ to 14$^{\circ}$ ), independent on the simulation domain size. This pitch angle is roughly consistent with current protoplanetary disk observations. We observed some weak trends regarding particle sizes and cooling time. On the other hand, considering the global nature of GI, it is desired to investigate both gas and dust spirals in global simulations in future. 

\acknowledgments
The authors thank the anonymous referee for valuable feedback. This research was supported by NASA TCAN award 80NSSC19K0639 and discussions with associated collaborators. Simulations made use of the Isaac cluster at the Max-Planck Center for Data and Computing in Garching and the Texas Advanced Computing Center (TACC) at the University of Texas at Austin through XSEDE grant TG-AST130002.

\software{Matplotlib \citep{Hunter2007}, SciPy \& NumPy \citep{Virtanen2020,vanderWalt2011}, IPython \citep{Perez2007}}

\bibliography{library}

\end{document}